# About Testing the Speed
# of Calculating the Shortest Route

**Assoc.Prof. Tiberiu Marius Karnyanszky, Ph.D.**
**Univ.Instr. Bogdan Ion Şelariu**
**„Tibiscus" University of Timişoara, Romania**

**ABSTRACT:** Applied into a various area of domains, the graph theory and its applications allow the determination of the shortest route. The common algorithm to solve this problem is Bellman-Kalaba, based on the matrix multiplying operation. If the graph is very large (e.g., the dimension of the associated incidence matrix is big), one of the main problems is to reduce the calculus time.
This paper presents a testing method able to analyze if an acceleration of the Bellman-Kalaba is possible and able to determine the time efficiency.

## 1.    Introduction

Calculating the shortest route in an orientated graph, in general, and in particular from the first to the last point (in the order they were numbered) can be done with the help of the Bellman-Kalaba algorithm presented in detail in [Kar05, Ion73, Kar04, Ren94].

## 2.    Material and method of calculus

A method of accelerating the calculus of the shortest route, described in [Kar05], is the one in which by multiplying the $A_m$ matrix with the $(v_n)^k$ column vector this being done from the last toward the first line, taking into consideration the new values obtained.

        To verify that the results obtained through this method are accurate and identical to those offered by the classical Bellman-Kalaba algorithm, we

have elaborated a program that carries out these comparisons, whit the following sequence of operations:

1) introduction of the initial data: number of points (between the limits $n_1$ and $n_2$), number of arks (between the limits $m_1$ and $m_2$), number of graphs to be generated;

2) generating the graphs and memorizing them in an external file. These graphs have, specific to each one, n points ($n_1 \leq n \leq n_2$) and m arcs ($m_1 \leq m \leq m_2$), where *n, m, i, j* and $a_{ij}$, *i=1,n, j=1,n*, have values generated randomly for each graph separately;

3) the data previously generated are read from the file and processed whit the classic Bellman-Kalaba algorithm, noting the total time needed for all the operations, for the entire set of graphs;

4) the process is repeated whit the modified Bellman-Kalaba algorithm, noting the time needed for the complete set of operations.

This comparative calculus program permits determinations on a set of graphs of any size, different from one another, generated by the computer, containing the following stochastic elements:

- each graph in particular can have a specific dimension, between the given limits $n_1$ and $n_2$; if $n_1 = n_2$ then the graphs will have the same dimensions;

- each graph in part can have a specific number of arcs, between the limits set by $m_1$ and $m_2$; if $m_1 = m_2$ then the graphs will have the same number of arcs. Although, the number of arcs respects the principle set through the program:

$$m \leq n*(n-1)/2 \qquad (1)$$

the case of equality being reached when the graph is complete, in other words it has all the possible arcs (the loops not being included);

- whatever the values of **n** and **m,** choosing the allocation of arcs in the graph (equivalent, the position of values in the adjacency matrix) is done randomly, the values of the pairs of points which comprise the arch (in the matrix, i and j) being different from matrix to matrix;

- the values associated with the arcs are random as well.

So then, the program, for a high number of graphs generated, can be considered, in conformity whit the principles of mathematical statistics, as acting upon a representative number of graphs and its results can be used to compare the speed of calculations of the two algorithms presented.

## 3.    Experimental results

To make a comparison between the two methods of determination, the classical one and the one whit the accelerated algorithm, we've made tests whit the sets of data from Tables 1 and 2 in which **n** represents the number of points, **m** represents the number of arcs, $t_{BK}$ represents the time of calculus for 1000 different graphs using the classic Bellman-Kalaba algorithm (in milliseconds), and $t_{BKaccelerat}$ represents the time of calculus for the same 1000 graphs using the accelerated Bellman-Kalaba algorithm (in milliseconds).

**Table 1. The comparison of calculus times for determinations with n and m of fixed values**

| n | m | $t_{BK}$ | $t_{BKaccelerat}$ | n | m | $t_{BK}$ | $t_{BKaccelerat}$ |
|---|---|---|---|---|---|---|---|
| 10 | 10 | 50 | 50 | 90 | 200 | 5110 | 4170 |
| | 30 | 50 | 50 | | 600 | 4440 | 3850 |
| | 50 | 50 | 50 | | 1000 | 4230 | 3790 |
| | 70 | 60 | 60 | | 1400 | 4120 | 3790 |
| | 90 | 111 | 110 | | 1800 | 4010 | 3740 |
| 30 | 10 | 390 | 380 | | 2200 | 3960 | 3620 |
| | 30 | 390 | 390 | | 2600 | 3900 | 3570 |
| | 50 | 500 | 490 | | 3000 | 3840 | 3520 |
| | 70 | 610 | 440 | | 3400 | 3790 | 3570 |
| | 90 | 550 | 490 | | 3800 | 3680 | 3510 |
| 50 | 10 | 890 | 940 | | 4200 | 3520 | 3520 |
| | 30 | 940 | 930 | | 4600 | 3510 | 3400 |
| | 50 | 1100 | 1050 | | 5000 | 3520 | 3460 |
| | 70 | 1380 | 1210 | | 5400 | 3460 | 3400 |
| | 90 | 1480 | 1320 | | 5800 | 3460 | 3400 |
| 70 | 10 | 1810 | 1700 | | 6200 | 3410 | 3300 |
| | 30 | 1810 | 1760 | | 6600 | 3400 | 3240 |
| | 50 | 1820 | 1860 | | 7000 | 3350 | 3190 |
| | 70 | 2140 | 1930 | | 7400 | 3300 | 3130 |
| | 90 | 2530 | 2140 | | 7800 | 3130 | 3080 |
| 90 | 10 | 2850 | 2850 | | | | |
| | 30 | 2860 | 2850 | | | | |
| | 50 | 3080 | 2910 | | | | |
| | 70 | 3300 | 3020 | | | | |
| | 90 | 3410 | 3190 | | | | |

**Table 2. The comparison of calculus times for determinations with n and m of interval type**

| n | m | $t_{BK}$ | $t_{BKaccelerat}$ |
|---|---|---|---|
| 10 – 30 | 001 – 100 | 270 | 220 |
| | 101 – 200 | 280 | 220 |
| | 201 – 300 | 220 | 170 |
| | 301 – 400 | 220 | 220 |
| | 401 – 500 | 220 | 220 |
| | 501 – 600 | 220 | 220 |
| | 601 – 700 | 220 | 220 |
| | 701 – 800 | 220 | 220 |
| 30 – 50 | 001 – 300 | 930 | 770 |
| | 301 – 600 | 830 | 770 |
| | 601 – 900 | 820 | 770 |
| | 901 – 1200 | 780 | 770 |
| | 1201 – 1500 | 780 | 770 |
| | 1501 – 1800 | 760 | 710 |
| | 1801 – 2100 | 710 | 710 |
| | 2101 – 2400 | 720 | 660 |
| 50 – 70 | 001 – 500 | 1980 | 1700 |
| | 501 – 1000 | 1920 | 1650 |
| | 1001 – 1500 | 1760 | 1700 |
| | 1501 – 2000 | 1700 | 1650 |
| | 2001 – 2501 | 1700 | 1540 |
| | 2501 – 3000 | 1590 | 1480 |
| | 3001 – 3500 | 1540 | 1480 |
| | 3501 – 4000 | 1490 | 1480 |
| | 4001 – 4500 | 1490 | 1480 |
| 70 – 90 | 0001 – 1000 | 3460 | 3020 |
| | 1001 – 2000 | 3180 | 2970 |
| | 2001 – 3000 | 3020 | 3850 |
| | 3001 – 4000 | 2810 | 2750 |
| | 4001 – 5000 | 2740 | 2690 |
| | 5001 – 6000 | 2690 | 2530 |
| | 6001 – 7000 | 2580 | 2470 |
| | 7001 – 8000 | 2530 | 2420 |

**Conclusions**

Analyzing the experimental results presented in Table 1, in which the number of points and arcs are fixed (predetermined), we can observe:

- As **n** increases, the time of calculus in both variants increases: for the same values of **m**(=10), the times of calculus are: $t_{BK, n=10}$=50 ms; $t_{BK, n=30}$=390 ms; $t_{BK, n=50}$=890 ms; $t_{BK, n=70}$=1810 ms; $t_{BK, n=90}$=2850 ms. The situation is normal, as the dimensions of the graph (and associated matrix) increase, the quantity of calculus is increased as well;

- For the same values of **n**, as **m** increases, but not approaching the maximum value, the time of calculus increases in both variants: $t_{BK, n=90, m=10}$=2850 ms; $t_{BK, n=90, m=30}$=2860 ms; $t_{BK, n=90, m=50}$=3080 ms; $t_{BK, n=90, m=70}$=3300 ms; $t_{BK, n=90, m=90}$=3410 ms. The situation is easily explainable, as the graph has more arcs, determining the shortest route involves more calculations;

- The increase presented at the previous point is nonlinear, it course harder at first then with larger differences (nothing that $t_{BK, n=90, m=10}$=2850 ms; $t_{BK, n=90, m=30}$=2860 ms, in other words for the increase from 10 to 30, with 20, of the number of arcs, the time increases just with 10 ms, while for $t_{BK, n=90, m=50}$=3080 ms; $t_{BK, n=90, m=70}$=3300 ms, meaning that for an increase of another 20 arcs the time increases by 210 ms). The explanation of this situation consists in the fact that if the number of arcs is small, the possibility of forming a road from $x_1$ to $x_n$ is also small and the algorithm finishes after the first two steps;

- When the values of **m** are large and approach the maximum value ( in conformity with (1)), the times for calculus drop: for example if n=90, when raising m (which can take values up to 90*89=8010) we can observe the time dropping from $t_{BK, n=90, m=1000}$=4230 ms to $t_{BK, n=90, m=3000}$=3840 ms and $t_{BK, n=90, m=5000}$=3520 ms or $t_{BK, n=90, m=7000}$=3350 ms. The situation is explainable through the fact that there are more and more possible routes to compare so it is easy to determine a minimum;

- In all the situations we can observe that $t_{BKaccelerat} \leq t_{BK}$. The equality is achieved rarely, only for small values of m, when, as we explained earlier, it is difficult to form a graph with a sufficient number of arcs to determine the rout from $x_1$ to $x_n$. The large difference, which means the reduction of the execution time, is at values between 10-15% at best.

Interpreting the data from Table 2, in which **n** and **m** are chosen randomly in a given interval, we can observe that:

- As the values of **m** increase, approaching the maximum given value in relation (1), the times of calculus drop: for example n=50-70, for the increase of m the time drops accordingly: for $BK, n=50\text{-}70, m=1001\text{-}1500$=1760 and $t_{BKaccelerat, \ n=50\text{-}70, \ m=1001\text{-}1500}$=1700 ms while for $t_{BK, \ n=50\text{-}70, \ m4001\text{-}4500}$=1490 ms and $t_{BKaccelerat, \ n=50\text{-}70, \ m=4001\text{-}4500}$=1480 ms;

- In all the situations we can observe that $t_{BKaccelerat} \leq t_{BK}$. The differences, translated in the reduction of execution times, are in this case as well between 10-15%.

So then, applying this program for generation and comparison of data, we can conclude that the modified calculus algorithm permits the reduction of the time for determining the value of the optimum route in an orientated graph.